\documentclass[runningheads]{llncs}
\usepackage{hyperref}
\usepackage{graphicx}
\usepackage{amsmath}
\begin{document}
\title{Statute-enhanced lexical retrieval of court cases for COLIEE 2022}
\author{Tobias Fink \and
Gabor Recski\and
Wojciech Kusa\and
Allan Hanbury}
\authorrunning{T. Fink et al.}
\institute{TU Wien, Faculty of Informatics, Research Unit E-commerce}
\maketitle              % typeset the header of the contribution
\begin{abstract}
We discuss our experiments for COLIEE Task 1, a court case retrieval competition using cases from the Federal Court of Canada. During experiments on the training data we observe that passage level retrieval with rank fusion outperforms document level retrieval. By explicitly adding extracted statute information to the queries and documents we can further improve the results. We submit two passage level runs to the competition, which achieve high recall but low precision.

\end{abstract}

\section{Introduction}

In the legal domain, court cases play an unique role as they often contain the last say on a particular legal subject. This is especially true in Common Law systems, such as the legal systems of North America, where court cases play a large role in shaping the law. While statutes are the foundation of the legal system, it is often necessary to look through precedent court cases for detailed information that is not available in statutes to reach a decision. However, not only are court cases long and difficult to read, the number of potentially relevant court cases is ever increasing. As such, the need for development of automated methods for retrieval of legal information to aid legal experts is equally increasing.

The Competition on Legal Information Extraction/Entailment (COLIEE)\footnote{\url{https://sites.ualberta.ca/~rabelo/COLIEE2022/}} evaluates legal information retrieval (IR) systems for a variety of legal retrieval tasks. We participate in the COLIEE 2022 Task 1, which deals with Canadian law precedent retrieval (notice cases). We experiment with lexical methods for retrieval, focusing on ways of improving established methods with domain-specific fine-tuning. Considering that statutes are still the foundation of the legal system, we add statute information to the search to focus the models on information that is typically defining relevancy in the legal domain. Although not all cases contain statute information, we observe that making use of this information will overall improve retrieval performance.

\section{Task Description}

In Task 1 of the COLIEE 2022, the goal is to retrieve supporting court cases (notice cases) for new court cases (query cases). Notice cases can be understood as precedent cases that are highly relevant for a query case. Each query case is supported by at least one notice case. For this task, cases from the Federal Court of Canada are used for both query and notice cases. A training collection as well as a test collection is provided (see Table \ref{tab:collections}), both having their own respective query cases, which are part of the collection. The training collection provides labels for relevant notice cases for each query, while the test collection only provides query cases without labels. Cases have been edited to have references to other cases removed and replaced by placeholder tokens. The task is to retrieve notice cases from the test collection using the queries of the test collection. The performance is measured using F1 score.

\begin{table}
\caption{Training and test collection statistics. Tokens per document and notice cases are per query.}\label{tab:collections}
\centering
\begin{tabular}{lcc}
\hline
 &  Training & Test\\
\hline
Total Cases &  4415 & 1563\\
Query Cases &  898 & 300\\
Max \# of tokens & 90567 & 61065\\
Median \# of tokens & 3658 & 3573\\
Mean \# of tokens & 4778 & 4979\\
Max \# of notice cases & 34 & N/A\\
Median \# of notice cases & 3 & N/A\\
Mean \# of notice cases & 4.68 & N/A\\
\end{tabular}
\end{table}

The length of the query documents makes the task challenging in a few ways. Not only are many IR methods better suited for shorter queries, due to the length of the documents, the relationship between query cases and notice cases is also difficult to understand without expert knowledge.

\section{Method}

We approach this task with the assumption that there is a topical overlap between query and notice cases, but that not all parts of a query case are equally important. It has been shown in past legal retrieval workshops (see AILA \cite{leburu-dingalo_ub_2020,liu_query_2020}, COLIEE \cite{rosa_yes_2021,ma_retrieving_2021,althammer_dossier_2021}) that lexical methods, such as BM25 or IR language models (LM), yield competitive results, even when compared to newer neural network based approaches. We build on top of these lexical methods and adapt them to the legal collections of this task.

\subsection{Document-Level} First, we experiment with using the models out-of-the-box. We preprocess the training collection by removing special characters and tokens with two characters or less and index the documents using Elasticsearch. Numbers are also removed, except when they are part of a statute section citation. During indexing, the text is also lowercased, stemmed and stopwords removed, including the task-specific placeholder tokens. To convert case documents to queries, we try to extract the most informative terms from the case. As a naive approach for this term extraction, we calculate the TF-IDF score for each token in the query case and then use the top $T$ tokens with the highest score as query terms. We compare the performance of the Elasticsearch implementations of BM25 and the LM Jelinek Mercer similarity \cite{zhai_study_2017}, which calculate a score $s$ for each document. For each query, 100 documents are retrieved and the precision, recall and F1 scores calculated for each rank. Although query cases are part of the collection, they are skipped during retrieval. We perform a random search to find the best hyperparameters for BM25 ($k1$, $b$) and LM Jelinek Mercer ($\lambda$) as well as $T$. While searching for the best hyperparameters, we only use the first 700 queries of the training collection (\textit{training set}). We determine the best cutoff rank $k$ using the F1 micro-averages for each rank. We use the remaining 198 queries (\textit{dev set}) to evaluate the best hyperparameters and cutoff rank $k$. 

\subsection{Passage-Level} Next, we experiment by changing the way how queries are created from query cases and change how documents are retrieved by using passage level retrieval. The information on where a passage starts and ends is already present in the case files and just needs to be utilized. Similar to the lexical baseline of \cite{althammer_parm_2022}, we split each case $c$ in the collection $C$ into passages $p_1,...,p_{n_c}$ and index the passages instead of the whole case, using the same preprocessing method as before. Now, the score $s$ is calculated for each passage instead of each document. A query case $q$ is also split into passage queries $pq_1,...,pq_{n_q}$ which retrieve a set of passage level rankings $R$ with $|R| = n_q$. We aggregate the passage level rankings to case level using Reciprocal Rank Fusion, a method of aggregation that outperforms other methods, such as Condorcet Fuse or CombMNZ \cite{cormack_reciprocal_2009}:

\begin{equation}
RRFscore(c \in C) = \sum_{r \in R} \frac{1}{k_{rrf} + r(c)} * p_b
\end{equation}

We set $k_{rrf} = 60$, the same value as in \cite{cormack_reciprocal_2009}. We also add a passage boost factor $p_b$ that is set to 1 for now. For this passage level ranking approach, we again perform the same method as before to find the best hyperparameters for BM25 ($k1$, $b$) and LM Jelinek Mercer ($\lambda$) as well as $T$ and $k$, using the same \textit{training set / dev set} split of queries for evaluation. For all further experiments, we the values of the hyperparameters are fixed to the best result of this random search (excluding cutoff rank $k$).

\subsection{Statute Field} Finally, we experiment with adding additional domain knowledge to the search by extracting statute sections mentioned in the case documents and adding them to the documents explicitly. For this purpose, we scrape the titles of Canadian rules, regulations, orders and acts from the Canadian Justice Law Website\footnote{\url{https://laws-lois.justice.gc.ca/eng/}}. This scraped list of titles also contains parts that would not be typically found in statute citations (e.g. text fragments that the law has been repealed). Consequently, we clean the titles by only considering text up to the first mention of \textit{regulations}, \textit{order}, \textit{act} or \textit{rules}. Further, since some statutes are only mentioned as acronyms, we create acronym candidates for each statute by taking the first upper-case letter of each token in the title. We identify the statutes of a case based on mentions of titles and generated title acronyms in the text. Additionally, we use regular expressions to detect statute section numbers in the text. We map statutes to section numbers by counting the number of passages in a case where a section number co-occurs with a statute mention, and then assigning the most frequently co-occurring statute to a section number.

These extracted statute sections are then added to the case passages and indexed as an additional statute-section field in Elasticsearch. We combine the original passage query with the extracted statute sections of the passage using a compound query. The score for the statute field $s_{statute}$ is calculated using BM25 and added to the overall score for each passage (the Elasticsearch default for a compound query), resulting in a new total score $s_{total}$:

\begin{equation}
s_{total} = s + s_{statute} * s_b
\end{equation}

To further control the influence of the statute-section field on the similarity calculation, we adjust the weight of the similarity score of the statute-section field with the factor $s_b$ (using the ElasticSearch \textit{boost} functionality). We assume that query passages that mention statute-sections are more likely to contain information that is of particular importance for a case. If the number of statute-sections $s_n$ that are present in the query passage is at least 1, we now set the earlier introduced passage boost factor $p_b$ to the hyperparameter $P_b$:

\begin{equation}
p_b = 
\begin{cases}
    P_b,& \text{if } s_n\geq 1\\
    1,              & \text{otherwise}
\end{cases}
\end{equation}

The best values for $P_b$ and $s_b$ are determined using a random search and $k$ is determined as before.

\section{Results}

\begin{table}
\caption{Results for our experiments using the \textit{training set} and \textit{dev set} queries.}\label{tab:results_training}
\centering
\begin{tabular}{lccc|ccc}

%Method & Training Set F1 & Dev Set F1\\
& \multicolumn{3}{c|}{Training Set} & \multicolumn{3}{c}{Dev Set}\\
\cline{2-7}
Method & Precision & Recall & F1 & Precision & Recall & F1\\
\hline
Document level BM25 ($k=7$) & 0.1193 & 0.1944 & 0.1479 & 0.0871 & 0.1825 & 0.1179\\
Document level LM ($k=8$) & 0.1200 & 0.2200 & 0.1553 & 0.0802 & 0.1892 & 0.1127\\
Passage level BM25 ($k=8$) & 0.1214 & \textbf{0.2226} & 0.1571 & 0.0898 & 0.2116 & 0.1261\\
Passage level LM ($k=8$) & 0.1210 & 0.2218 & 0.1565 & 0.0993 & \textbf{0.2341} & 0.1395\\
Passage level LM + Statute Field ($k=7$) & \textbf{0.1282} & 0.2090 & \textbf{0.1589} & \textbf{0.1073} & 0.2249 & \textbf{0.1453}\\
\end{tabular}
\end{table}

\begin{table}
\caption{Excerpt of the task 1 ranking showing selected runs and our results.}\label{tab:results_runs}
\centering
\begin{tabular}{lllccc}
\hline
Rank & Team & Run & F1 Score & Precision & Recall\\
\hline
1 & UA & pp\_0.65\_10\_3.csv & \textbf{0.3715} & 0.4111 & 0.3389\\
2 & UA & pp\_0.7\_9\_2.csv & 0.3710 & \textbf{0.4967} & 0.2961\\
3 & siat & siatrun1.txt & 0.3691 & 0.3005 & \textbf{0.4782}\\
7 & LeiBi & run\_bm25.txt & 0.2923 & 0.3000 & 0.2850\\
15 & TUWBR & TUWBR\_LM\_law & 0.2367 & 0.1895 & 0.3151\\
17 & TUWBR & TUWBR\_LM & 0.2206 & 0.1683 & 0.3199\\
\end{tabular}
\end{table}

The results of the experiments on the training set and dev set are shown in Table \ref{tab:results_training}. On \textit{document level}, \textit{BM25} achieved the highest F1 using the parameters $T = 200$, $k1 ~= 1.09$, $b ~= 0.99$ while the \textit{LM} Jelinek Mercer achieved the highest F1 using $T = 200$, $\lambda ~= 0.64$, $b ~= 0.99$. On \textit{passage level}, \textit{BM25} achieved the highest F1 using the parameters $T = 100$, $k1 ~= 0.66$, $b ~= 0.59$ while the \textit{LM} Jelinek Mercer achieved the highest F1 without limiting $T$ and $\lambda ~= 0.56$. This means \textit{LM} uses every token of a passage as query, but with duplicate tokens removed. In our experiments, all passage level methods with rank fusion outperform document level retrieval methods. For this reason we did not continue experimenting on document level. While passage level BM25 achieved a higher F1 score on our training set, the LM model performed better on the dev set. The best overall F1 score was achieved by the LM model with inclusion of the statute field. Especially the dev set performance could be improved by adding this information.

For the task submission, we submitted two runs, using the \textit{Passage level LM} setup as run \textbf{TUWBR\_LM} and using the \textit{Passage level LM + Statute Field} setup as run \textbf{TUWBR\_LM\_LAW}. The results for our submitted runs and a selection of top scoring runs for the task are shown in Table \ref{tab:results_runs}. Our methods achieve a high level of recall but perform poorly regarding precision. However, our runs are only situated in the bottom half of the F1 score sorted ranking.

One weakness of our method is certainly that our naive term extraction approach was insufficient. Further, we were only able to produce a ranking of court cases and determined relevancy based on a fixed cutoff value (rank 7 or 8). Since most query cases cite fewer notice cases than our cutoff value, our precision is low. However, extracting statute-section information produced positive results. If we compare our two runs, we can see that adding statute information can yield a higher precision while recall is only reduced minimally. We expect that results can be improved further with better strategies for utilizing this information.

\section{Conclusion}

For the COLIEE 2022 Task 1 case retrieval, we used passage-level LMs to retrieve notice cases for case queries. Our methods achieved a high recall but low precision. We showed that a simple method making use of statute-section mentions in passages can achieve a higher precision with only a minor decrease in recall. Overall, low precision remained a problem for our methods and they were outperformed by other methods in the competition. We expect that our lexical approach could still be improved by different query term extraction strategies.\\

\noindent
\textbf{Acknowledgments.}
Project partly supported by BRISE-Vienna (UIA04-081), a European Union Urban
Innovative Actions project.

\bibliographystyle{splncs04}
\bibliography{Zotero}
\end{document}